\documentclass[published]{JHEP3} 
\usepackage{epsfig}
\epsfclipon

\newcommand{\nco}{\newcommand}
\nco{\beq}{\begin{equation}}
\nco{\eeq}{\end{equation}}
\nco{\beqa}{\begin{eqnarray}}
\nco{\eeqa}{\end{eqnarray}}
\nco{\lra}{\leftrightarrow}

\def\sgn{{\rm sgn}}
\nco{\sss}{\scriptscriptstyle}
\nco{\vpp}{\vec p^{\,2}} 
\nco{\lsim}{\mbox{\raisebox{-.6ex}{~$\stackrel{<}{\sim}$~}}}
\nco{\gsim}{\mbox{\raisebox{-.6ex}{~$\stackrel{>}{\sim}$~}}}

\title{\Large Reheating from Tachyon Condensation}

\author{James M.~Cline, Hassan Firouzjahi, Patrick Martineau\\
Physics Department, McGill University,
3600 University Street, Montr\'eal, Qu\'ebec, Canada H3A 2T8\\
E-mail: \email{jcline@physics.mcgill.ca}, \email{firouzh@physics.mcgill.ca},
\email{martineau@physics.mcgill.ca}}
\preprint{McGill 02-21}
\received{\today}              
\accepted{\today}              
\JHEP{12(2001)999}             

\abstract{
We argue that it may be possible to reheat the universe after inflation driven by D-brane
annihilation, due to the coupling of massless fields to the time-dependent tachyon condensate
which describes the annihilation process.  This mechanism can work if the original branes
annihilate to a stable brane containing the standard model.  Given reasonable assumptions about
the shape of the tachyon background configuration and the size of the relevant extra
dimension,  the reheating can be efficient enough to overcome the problem of the universe being
perpetually dominated by cold dark tachyon matter.  In particular, reheating is most efficient
when the final brane codimension is large, and when the derivatives of the tachyon
background are large.}

\begin{document}
\section{Introduction}

Because of the difficulty of finding direct laboratory probes of string theory, it is
interesting to look for possible evidence from cosmology.   Most notably, inflation may be
tested more sensitively in the near future by the MAP \cite{map} and PLANCK \cite{planck}
observations of the cosmic microwave background radiation.  Recently there has been significant
progress in constructing stringy inflation models which make use of naturally occuring
potentials between D-branes to provide the false vacuum energy \cite{inf,ST,JST}.   However, these
attempts have not adequately addressed the question of how reheating occurs after inflation. 
In fact, there are reasons to fear that reheating may be generically difficult to achieve in
D-brane inflation.  For this reason, we aim to propose a generic mechanism for reheating in
such models, which is qualitatively different from reheating in ordinary field-theory inflation
models, and which has the hope of being fairly robust. It is based on particle production in a
time-varying background,  which will occur even if the background motion is not
oscillatory.

Let us begin by describing the difficulty with reheating. In the simplest version of D-brane
inflation, a parallel brane and antibrane begin with some separation between them in one of the
extra dimensions required by string theory.  Although parallel branes are supersymmetric and
have no force between them, the brane-antibrane system breaks supersymmetry so that there is an
attractive force and hence a nonvanishing potential energy.  It is the latter which drives
inflation.  Once the branes have reached a critical separation, they become unstable to
annihilation.  The instability is described by condensation of a tachyonic mode \cite{sen}. 
Its low energy effective description is a field
theory  of a peculiar kind, whose Lagrangian has the form \cite{BSFT1,BSFT2}\footnote{we
use the metric signature  $(1,-1,\dots,-1)$}
\beq
\label{action1}
  {\cal L} = -{\cal T} e^{-T^2/a^2} F[-(\partial_\mu T)^2]
\eeq
where 
\beq
\label{Feq}
	F(x) = {4^x x \Gamma(x)^2\over 2\Gamma(2x)},
\eeq
which is determined for the superstring from the boundary string field theory
(BSFT). 
Here ${\cal T}$ is the sum of the original brane tensions and $a$ is of order the
string length, $a\sim\sqrt{\alpha'}$ (the precise value to be used in the model
we adopt will be discussed in section
IV).  The tachyon starts from the unstable maximum
$T=0$ and rolls to $T\to\infty$.  This process requires an infinite amount of time,
during which the tachyon fluid has an equation of state identical to that of pressureless
dust as $\dot T\to 1$ \cite{sen_tm,tach_fluid}.  

(There have been numerous recent attempts to make use of the tachyon fluid for cosmology
\cite{tach_cosm},
either as the inflaton or as quintessence.  Although these ideas might work if one had
the freedom to change the form of the tachyon potential, the action which arises from
string theory is not suitable for either purpose.  Since the late-time equation of state is
$p=0$, the string theory tachyon does not provide accelerated expansion in the recent universe.  At 
early times, its potential is too steep to satisfy the requirements of inflation.
Constraints on tachyon cosmology and its shortcomings have been discussed in
\cite{SW} and \cite{KL}.)

Returning to our description of the endpoint of D-brane inflation, the situation is
similar to that of hybrid inflation \cite{linde}, where the tachyon plays the role of the unstable
direction in field space which allows for inflation to quickly end.  The important
difference is that in a normal hybrid inflation model, $T$ would have a minimum at some
finite value, {\it e.g.,} due to a potential like  $\lambda(|T|^2-a^2)^2$, and the
oscillations of $T$ around its minimum could give rise to reheating in the usual way, or the
more efficient tachyonic preheating \cite{preheat}.
But with the exponential potential in (\ref{action1}) there can be no such oscillations. 
It thus appears that the universe will become immediately dominated by the cold tachyon
fluid, and never resemble the big bang \cite{KL}.  It is important to convert very nearly
100\% of the energy stored in tachyon matter into radiation because the former redshifts
more slowly than the latter.  For example if we assume that the universe started initially
with a temperature of at least 1 TeV for the purposes of baryogenesis, while matter
domination begins at a temperature near 0.1 eV, then the ratio of energy density in
tachyon matter versus that in radiation must have been no more than $10^{-13}$.

To emphasize the difficulty of getting such efficient reheating in the present
situation, let us contrast the rolling tachyon with an inflaton which
is decaying through its oscillations into massless fermions via a coupling 
$g\phi\bar\psi\psi$.  The probability of particle production in the background
$\phi = \phi_0\sin(Mt)$ is proportional to the square of the transition matrix element
$\int dt \sin(Mt) e^{2i\omega t}\sim \delta(M-2\omega)$, where $\omega$ is the energy of the 
fermion or antifermion.  The square of the delta function is understood in the usual 
fashion to be $\delta(M-2\omega)$ times the total time; in other words, oscillations lead to a
constant {\it rate} of particle production.  By simply waiting long enough, the energy stored
in the oscillations will naturally be reduced to an exponentially small level.  In the
case of the rolling tachyon however, $\phi$ is replaced by $T$ which has the asymptotic
behavior $T\sim t$, and the whole action is multiplied by a factor $e^{-T/a}$ or
$e^{-T^2/a^2}$.  The corresponding transition matrix element
$\int dt  t e^{-t/a+ 2i\omega t}$ is finite, with no delta function.  As a result, it gives
not a constant rate of particle production, but rather a finite number density of
produced particles.   It is not obvious that the energy stored in the tachyon fluid can
be sufficiently reduced.

Nevertheless, it is known that the tachyon couples to massless gauge fields;
one form that has been suggested for the low energy theory is the Dirac-Born-Infeld
action \cite{action,sen_action},\footnote{This form is not the same as what one derives from
BSFT. However it has the right qualitative features, besides
being simpler to work with.  Ignoring for a moment the gauge field, it 
can be shown that $\sqrt{|\det(g_{\mu\nu} - \partial_\mu T\partial_\nu T )|}=
\sqrt{1-\dot T^2}$ for time-dependent configurations, so that the action vanishes
as $\dot T\to 1$.  The same is true for the BSFT-derived function $F(-\dot T^2)$
in (\ref{Feq}).  Moreover both functions behave like $\sqrt{(\partial_x T)^2}$ for
spatially dependent profiles in the limit that $\partial_x T\to\infty$, whose relevance
will be explained in the next section.}
\beq
\label{action2}
  {\cal L} = -{\cal T} e^{-T^2/a^2}
  \sqrt{|\det(g_{\mu\nu} + F_{\mu\nu} - \partial_\mu T\partial_\nu T )|}
\eeq
Because of its time dependence, we expect that some radiation will be produced by the
rolling of the tachyon.  It then becomes a quantitative question: can this effect be
efficient enough to strongly deplete the energy density of the tachyon fluid, so that the
universe starts out being dominated by radiation rather than cold dark matter?  There is
one immediate problem with this idea, however; the fact that the entire action is
multiplied by the factor $e^{-T^2/a^2}$ means that the massless particles  which are produced
will not act like ordinary radiation \cite{tach_gauge1}.  Recent work has shown that these
excitations have the same equation of state as the tachyon itself \cite{tach_gauge2}.  
This is related to the fact that the system is annihilating to the closed string vacuum,
which does not support any open string states like the gauge fields.  

An obvious way to circumvent the above difficulty is to assume that the branes  annihilate
not to the vacuum, but rather to branes of lower dimension  \cite{sen_descent}. This is a
natural possibility since any initial fluctuations of the tachyon field where it changes sign
will lead to the production of topological defects at the points where $T=0$, via the Kibble
mechanism \cite{ST}.  In the effective description, the lower dimensional branes are
represented by solitonic configurations of the tachyon field.  A $D(p-2)$ brane is a vortex
of the  complex tachyon field, whereas a $D(p-1)$ brane is a kink \cite{BSFT1,BSFT2,MZ} .
These possibilities are  described by the descent relations of Sen \cite{sen_descent}.  The
stable descendant branes are able to support open string excitations of gauge fields,
including those of the standard model.   They will no longer decouple as a consequence of the
rolling tachyon because the  topological defect which prevents pins $T=0$ at the origin.

In this paper we will set up a simplified model of particle production by tachyon
condensation, which we hope captures the essential features of a more realistic Lagangian.
The computational method developed here should carry over straightforwardly to more
complicated situations.  In section II we motivate an ansatz for the space- and
time-dependent tachyon background which describes condensation to a brane of one  dimension
lower than the initial configuration.  In section III we present the solutions for a gauge
field in this classical tachyon background.  Section IV describes how we compute the
spectrum of particles produced during the early phase of the tachyon motion, and
presents numerical results.  We conclude in section V with the interpretation of these
results and a discussion of how the calculation can be improved in future work.

\section{Tachyon background}

There are two kinds of tachyonic solutions which have been described in the literature: (1)
static solutions which are topological defects and represent lower dimensional branes, and
(2) dynamical solutions which are spatially homogeneous and describe a cosmological fluid
with vanishing pressure at late times.  The mechanism we have in mind combines these two
pictures by supposing that at $t=0$ the tachyon starts from (or very near to) the unstable
equilibrium configuration $T=0$ throughout some number $d$ of extra dimensions which will be
transverse to the final descendant brane.  We denote the spatial coordinates of these extra
dimensions by $x_1,\dots,x_d$.  Starting at $t=0$, the tachyon starts rolling towards
$T\to\infty$ for $|x_i|$ sufficiently far from the center, whereas it is pinned to $T=0$ at
$x_i=0$.  By charge conservation, there must be at least one other position where $T=0$
also.  The pair of defects represents a brane and an antibrane.  For simplicity we will focus
our attention on the half of the internal space near $x_i=0$ which contains the brane, since
by symmetry the physics near the antibrane  will be identical, for the purposes of particle
production. At any given time, this half of the extra dimensional space  will contain an
interior region where the defect resides, and an exterior one where the tachyon is still
rolling.  We might expect the tachyon profile to evolve with time similarly to fig.\ 1(a).

\medskip
\centerline{\epsfxsize=2.5in\epsfbox{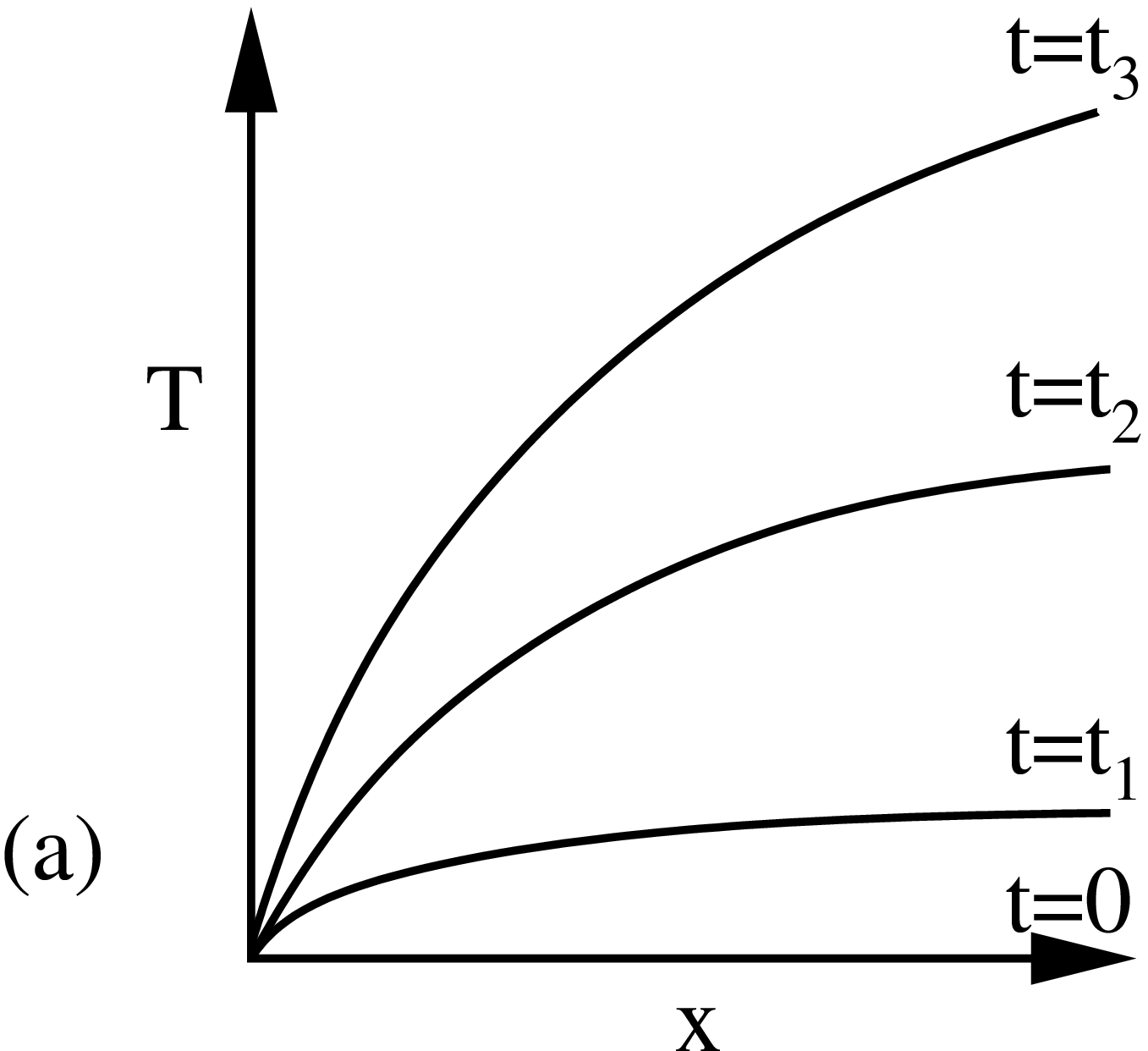}\hfil\epsfxsize=3.0in\epsfbox{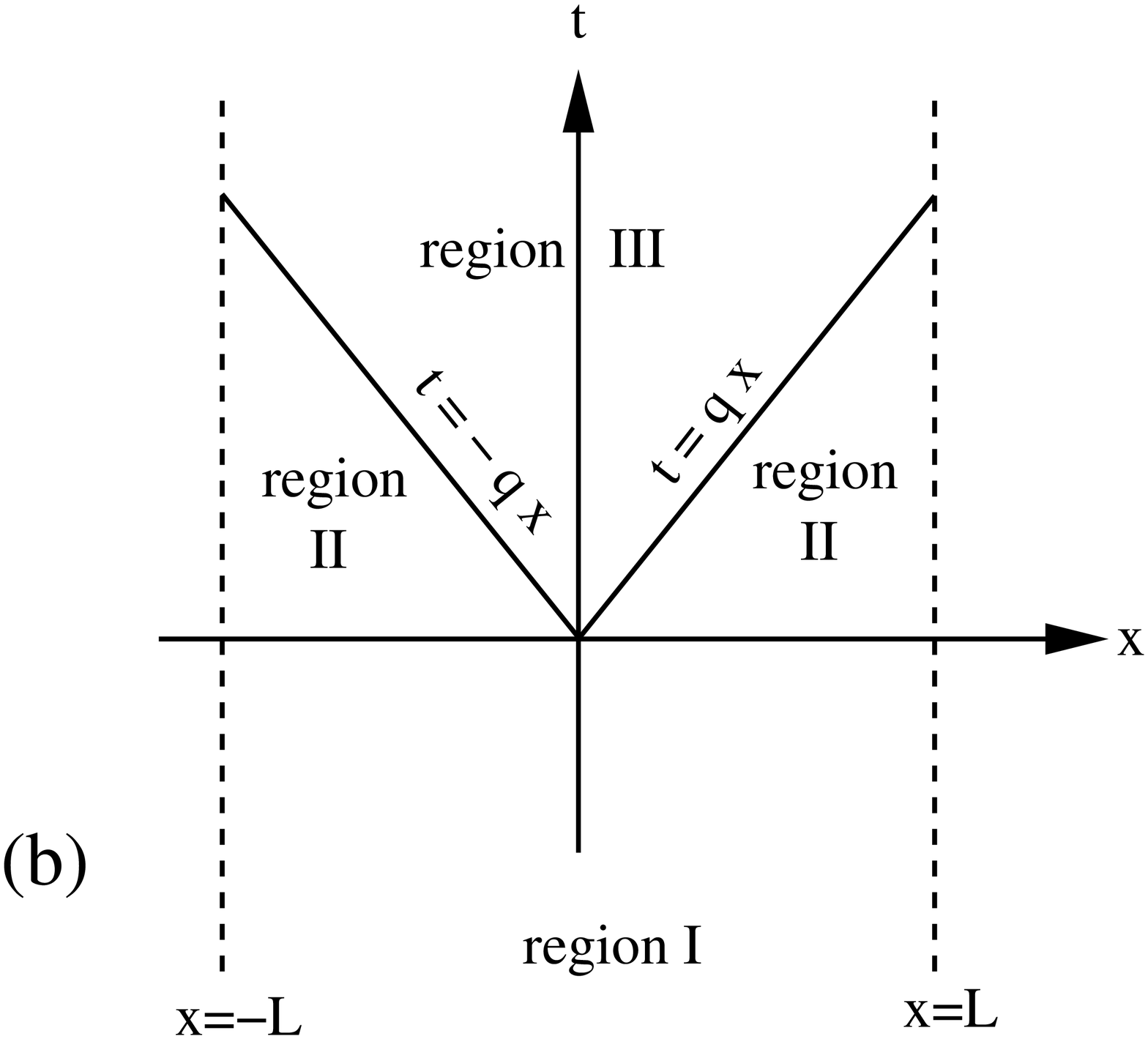}}
{\small\noindent
Figure 1. (a) Possible time evolution of the tachyon field $T$ during condensation.
(b) Our ansatz for the tachyon evolution in time and in the half of the extra dimensional
space containing the defect representing the final state brane: region III contains a linear
kink, II contains the homogeneous rolling field, and I is the unstable vacuum
configuration.}
\medskip

To find the real behavior of $T(t,x_i)$ we should solve the equation of motion for
$T$ which follows from eq.\ (\ref{action1}).  This is already a difficult numerical task
in itself, and its results would surely complicate our next step, which will be to
solve the gauge field equation of motion in the background $T(t,x_i)$.  Therefore
we are going to satisfy ourselves with a simplified ansatz for the background and
defer more detailed investigations to future work.  Namely, we will consider a linear
profile for the defect, and a purely time-dependent one in the exterior:
\beqa
\label{RIII}
	T(x) &=& qx,\qquad\qquad\qquad\qquad\ \ \, |x| < t/q \qquad(\hbox{region III})\\
\label{RII}
	T(t,x) &=& a\ln(\cosh(t/a))\sgn(x), \quad\! |x|> t/q \qquad\!(\hbox{region II})\\
\label{RI}
	T &=& 0, \qquad\qquad\qquad\qquad\qquad t<0 \qquad\ \ \ (\hbox{region I})
\eeqa	
We have specialized to the case of a single codimension for the descendant brane ($p=1$) and
this dimension has been compactified by identifying the points $x=2L$ and $x=-2L$, where an
antikink is assumed to reside.  We show only the half of the space $-L<x<L$ which surrounds
the kink. A parameter $q$ describing the steepness of the tachyon kink profile has been 
introduced.  The spacetime regions I-III are illustrated in figure 1(b).

One may wonder to what extent the ansatz (\ref{RIII}-\ref{RI}) reproduces a more
realistic tachyon background.  Let us first consider the rolling tachyon region (II).  
Eq.\ (\ref{RII}) is in fact an exact solution not for the action (\ref{action1}) but
rather for a mutilated version in which the argument of the exponential is linear:
\beq
\label{action3}
 {\cal L} = -{\cal T} 
 e^{-|T|/{a}}\sqrt{|\det(g_{\mu\nu}- \partial_\mu T\partial_\nu T )|}
\eeq
This closely resembles the BSFT result for the effective action from the bosonic string,
except for having $|T|$ rather than $T$ in the exponent.  The bosonic string has an
unphysical instability as $T\to -\infty$ which we artificially remove by taking the absolute value.
The behavior of the solutions to (\ref{action3}) is similar to that coming from 
a potential with a quadratic argument, which is the appropriate one for the
superstring.  For example, the late-time behavior corresponding
to the Lagrangian ${\cal L} = {\cal T} e^{-T^2/a^2}(1-\dot T^2)^b$ is 
\beq
\label{actualT}
	T(t) \cong t + {a^2(1-b)\over 4 t}(2b)^{1/(1-b)} e^{-t^2/(a^2(1-b))} + O(e^{-2t^2/(a^2(1-b))})
\eeq
assuming that $b<1$.  (This can most easily be derived by computing the corresponding
Hamiltonian density and using energy conservation to get a first integral of the
motion.)  Similarly, if we expand (\ref{RII}) for large times, we get
\beq
	T(t) \cong t + a e^{-2t/a} + O(e^{-4t/a})
\eeq
which resembles (\ref{actualT}).
Moreover if we compute
the relevant factor $\sqrt{1-\dot T^2}$ in the two cases (now taking $b=1/2$),
\beq
\sqrt{1-\dot T^2} \quad = \quad e^{-t^2/a^2} + O(e^{-2t^2/a^2}), \quad e^{-t/a} + O(e^{-2t/a})
\eeq
respectively.  In other words $\sqrt{1-\dot T^2}$ is just equal to the other prefactor
$e^{-t^2/a^2}$ or $e^{-t/a}$, as the case may be. This can easily be understood from the
fact that the conserved Hamiltonian is $V(T)(1-\dot T^2)^{-1/2}$, hence $V(T)$ and
$\sqrt{1-\dot T^2}$ must be proportional to each other. It will turn out that the
simplified form $e^{-t/a}$ makes it easier to solve for the gauge field in the next
section, which is one of our motivations for adopting the action (\ref{action3}).

Next we discuss the ansatz for the kink profile, $T=qx$, in region III.  This static
profile is not a solution to the equations of motion\footnote{an exact solution can be
obtained from that of region II by analytically continuing $t\to ix$, giving  \beq T = a
\,\sgn(x) \ln\left({\cos((L-x)/a)\over\cos(L/a)}\right),\nonumber\eeq  where we have
imposed continuity of $T'$ at $x=\pm L$ and assumed that $L/a\le \pi/2$.  The solutions for
the gauge field are complicated in this background, so we prefer to use the linear kink
profile for this paper.} except in the limit that $q\to\infty$.  If we consider a
generalized action  of the form ${\cal L} = V(T)(1-T'^2)^b$, then the linear profile is
a solution only if $q^2=1/(2b-1)$.  The limit $q\to\infty$ has been discussed in
\cite{BSFT1,BSFT2}, where it was noted that the descendant brane resulting from tachyon
condensation has the right tension to  agree with string theory when $q\to\infty$.  In this
limit the descendant brane looks like a genuinely lower dimensional object, whereas it
would have a nonvanishing thickness (revealed by the energy density of the profile) for
finite values of $q$.   However a different method, that of level-truncation
\cite{LT1,LT2}, leads to tachyon defect  profiles which do have a nonzero width.  In any
case, there is no reason to believe that the kink thickness goes to zero immediately; it
seems quite reasonable to assume that it will have some nonzero value initially, and
possibly tend toward zero only as $t\to\infty$. We will assume that $q$ remains 
approximately constant for $0<t<qL$ since this is the time during which particles are
produced.  If $q\to\infty$ at later times, it will not significantly change our conclusions.

A further issue is the shape of the spacetime boundary separating the static kink from
the time-dependent solution. We have taken it to be linear, $t = q|x|$, but this implies
that $T$ is not even continuous at the interface except for $t\gg a$.  It might seem more
reasonable to deform the boundary such that $e^{q|x|/a} = e^{t/a}+e^{-t/a}$. We could do
so, but it would needlessly complicate the ansatz without even solving the problem of the
tachyon Lagrangian being discontinuous at the interface, because the factor
$\sqrt{|\det(g_{\mu\nu} - T_{\mu\nu})|} = \sqrt{1-(\partial_\mu T)^2}$ is still not
smooth:  $(\partial_\mu T)^2$ changes sign as well as magnitude across the interface:
\beq
(\partial_\mu T)^2 \cong \left\{\begin{array}{ll} 1- e^{-2t/a},& \hbox{region II}\\
						-q^2, & \hbox{region III}\end{array}\right.
\eeq
This is an indication that our ansatz (\ref{RIII}-\ref{RI}) is defective; the true
solution should have a smooth Lagrangian density at the transition.  Nevertheless, it
would greatly complicate the solution for the gauge fields in the next section if we
tried to smooth out this behavior.  We will instead compensate for the difficulties which
arise from this oversimplification in another way, as will be explained.

We found it simpler to consider condensation to a kink instead of a vortex
configuration of the tachyon.  The latter would be more realistic because of the fact
that, assuming the parent branes are supersymmetric BPS states, a brane of only one
dimension less is not, and consequently not stable, whereas a brane whose dimensionality
differs by an even number {\it is} BPS.  To reduce the dimensionality by two, the 
tachyon should be a complex field which condenses to a codimension-two defect, a vortex.
Although we originally wanted to treat this case, it is not clear how to write the 
action of the complex tachyon in a way which extends to the time-dependent configurations
of region II.   BSFT calculations show that for a winding configuration of the form 
\beq
	T = q_1 x_1 + i q_2 x_2
\eeq
the Lagrangian factorizes as \cite{BSFT2}
\beq
\label{action4}
	{\cal L} \propto -F(q_1^2) F(q_2^2)
\eeq
where $F(x) = {4^x x \Gamma(x)^2\over 2\Gamma(2x)}$.   It is not immediately obvious how to
rewrite (\ref{action4}) in a Lorentz and  gauge invariant way which would allow us to
deduce the equations of motion for time-dependent homogeneous configuations.  Explicit
constructions involve the use of independent tensors like $\partial_\mu T^* \partial_\nu T$
and $\partial_\mu T \partial_\nu T$ in matching powers of derivatives, so that a compact
expression like (\ref{action3})  does not seem to emerge.  We leave the consideration of
these complications for future work.

That being said, the kink configuration may still be physically relevant because
of the possibility of descending from the original $Dp$-$\overline{Dp}$ pair in two steps,
with the unstable $D(p-1)$ brane being a resonance through which the system passes on
its way to the stable $D(p-2)$ endpoint \cite{HN}.  The transition from $D(p-1)$ to $D(p-2)$ would
be described by the formation of a kink.

Finally, let us consider the fact that the rolling phase of the tachyon field
ends in a finite time within our ansatz. Fig.\
1(b) shows that at late times we have eliminated the homogeneous condensate by fiat
since at $t=qL$ the entire bulk has been replaced by the static kink.  We don't know
whether this is the actual behavior, or if the homogeneous region 
persists, which could be the case if $q$ grows with time:
\beqa
\label{RIIIa}
	T(t,x) &=& q(t)|x|,\qquad\quad\quad\qquad\phantom{\sgn(x)}
  |x| < t/q(t) \quad(\hbox{region III})\\
\label{RIIa}
	T(t,x) &=& a\ln(\cosh(t/a))\sgn(x), \qquad\ |x|> t/q(t) \quad(\hbox{region II})
\eeqa
If $q(t)$ grows with $t$ faster than linearly, then region II survives and the tachyon
fluid coexists with the final state brane at arbitrarily late times.  We believe that
the present calculation could give a reasonable approximation to the efficiency of
particle production even in this case.  The compactification length $L$ will be replaced
by the size of region III at the characteristic time scale when the fast roll phase ends
({\it i.e.} when $\dot T\cong 1$ in the bulk). On the other hand,  if $q(t)/t\to 0$ as
$t\to\infty$, then as long as the bulk is compact, region II  disappears completely.  In
this case it is still important to consider particle production on the brane since
otherwise the energy that was stored in the rolling tachyon might go into invisible
closed string modes in the bulk, namely gravitons, which would not be an acceptable
form of reheating.

\section{Gauge field solutions}

Our aim is to find out whether the energy stored in the homogeneous tachyon fluid can be
efficiently converted into radiation, so that the universe at least has a long period of
radiation domination before possibly giving way to the cold dark matter of the rolling
tachyon condensate.  We will do this by quantizing the gauge field in the tachyon 
background and computing the Bogoliubov coefficients that quantify the mismatch 
between the vacuum states of regions I and III (see fig.\ 1(b)); see for 
example \cite{BD,PV}.  That is, if we start in the vacuum
state appropriate for region I, we find that it is no longer the vacuum in region III,
and therefore radiation must be produced.  

The first step is to find the action for the gauge fields to quadratic order in the
fields.  Expanding (\ref{action3}) in the tachyon background described in the previous
section, we obtain
\beq
\label{action}
	S = \frac12 {\cal T}\int dt\, dx\, d^{\,3}y \left\{
		\begin{array}{ll} ({\partial_t\vec A})^{2} - (\nabla_y \vec A)^2 - 
			(\partial_x\vec {A})^2, & \hbox{region I}\\
 ({\partial_t\vec A})^{2} - e^{-2 t/a}\left( (\nabla_y \vec A)^2 + 
			(\partial_x\vec {A})^2 \right),& \hbox{region II}\\
\sqrt{1+q^2}\, e^{-q |x|/a} \left( ({\partial_t\vec A})^{2} - (\nabla_y \vec A)^2
	-{1\over 1+q^2}(\partial_x\vec {A})^2 \right), &  \hbox{region III}
			\end{array} \right.
\eeq
where $y_i$ are the coordinates of the large 3 dimensions, and we have absorbed the
volume of any other compact dimensions which are merely spectators into the brane tension
${\cal T}$.  We employed radiation gauge ($A_0=\nabla_y\cdot \vec A=0$) and projected out
the extra polarization by setting $A_x=0$, which is consistent since this state turns out
to have a mass gap in region III, unlike the massless components among the large three
dimensions, $\vec A$.  (The factor $e^{-2 t/a}$ in region II should really be
$(\cosh(t/a))^{-2}$, but this not an important difference, in the spirit of the other
approximations we have made.)  In the following we will drop the polarization indices
of the gauge field and write simply $A$ instead of $\vec A$.

To make our analysis more tractable, we have ignored the time dependence of the 
metric due to the expansion of the universe in the above action.  This neglect 
can be justified if the initial fast-roll regime of the tachyon, during which most of
the particle production occurs, does not take more than approximately one Hubble time.
We expect that this will be true if the string scale is somewhat below the Planck
scale and strings are weakly coupled, since then $H \sim \sqrt{2L{\cal T}}/
M_p^2\sim g_s M_s/(2\pi M_p)$ [using the relation $g_s^2M_p^2 = M_s^8 V/\pi(2^\pi)^6$
\cite{JST} and eq.\ (\ref{tension})],
whereas the time scale for the tachyon roll is of order the string scale. 

In the previous section we gave detailed motivations for our choice of the ansatz for
the classical tachyon background.  It is worth emphasizing one further criterion:
by choosing $T(t,x)$ to depend only on $t$ or $x$ in each region, we insure that the
gauge field equations of motion can be solved using separation of variables, which
greatly simplifies the task.

\subsection{Solutions in each region}

The solutions in region I are trivial since the background tachyon configuration is
simply $T=0$:
\beqa
	A_{\sss I} ={1\over\sqrt{4L}}\sum_m {1\over\sqrt{2\omega_m}}\Bigl[
	(a_m e^{-i\omega_m t} + a^\dagger_m e^{i\omega_m t}) \cos(k_m x) 
	+ (\tilde a_m e^{-i\omega_m t} + \tilde a^\dagger_m e^{i\omega_m t}) \sin(k_m x)
	\Bigr]\nonumber\\
\eeqa
where $k_m = m\pi/L$,  $\omega_m^2 = \vpp + k_m^2$, and $\vec p$ is the momentum in
the three large dimensions.  We have split the modes according to their parity in the
extra dimension for later convenience.

In region II, the equation of motion for $A$ is
\beq
	\ddot A + \omega_m^2 e^{-2t/a} A = 0,
\eeq
which has the solutions
\beqa
   A_{\sss II} &=& {1\over\sqrt{4L}}\sum_m \Bigl[
   \left(b_m J_0(a\omega_m e^{-t/a}) + c_m Y_0(a\omega_m e^{-t/a})\right)
   \cos(k_m x) \nonumber\\
	&& + \hbox{\ same with\ } b_m \to \tilde b_m,\  c_m \to \tilde c_m,\ 
		\cos(k_m x)\to\sin(k_m x)	\Bigr]
\eeqa
Near $t=0$, these oscillate just like the region I solutions, but at large $t$ the
oscillations freeze and the solutions grow linearly with time.

Region III is the important one at late times, since this is where the descendant 
brane and the standard model are supposed to reside.  Here the equation of motion is 
\beq
	(1+q^2)\left(-\ddot A + \nabla^2_y A\right) + A'' - {q\over a}\sgn(x) A' = 0
\eeq
using primes to denote $\partial_x$.  The solutions can be written as
\beqa
	A_{\sss III} =\sum_n {1\over\sqrt{2\bar\omega_n}}\Bigl[
	(d_n e^{-i\bar\omega_n t} + d^\dagger_n e^{i\bar\omega_n t}) f_n(x) 
	+ (\tilde d_n e^{-i\bar\omega_n t} + \tilde d^\dagger_n e^{i\bar\omega_n t})
	\tilde f_n(x)
	\Bigr]
\eeqa
with
\beqa
\label{baromega}
	\bar\omega_n &=& \left\{\begin{array}{ll} \sqrt{\vpp}, & \phantom{AAAAAAA} n=0\\
		\left[ \vpp + {1\over 1+q^2}\left( {q^2\over 4 a^2} + k_n^2\right)
			\right]^{1/2},& \phantom{AAAAAAA} n\ge 1 \end{array} \right. \nonumber\\
	f_n(x) &=&\left\{\begin{array}{ll} 
		N_0, & n=0\\
		 N_n e^{q|x|/2a}\left(\cos(k_n x) - {q\over 2 k_n a}\sin(k_n|x|)\right)	
			,& n\ge 1 \end{array} \right. \nonumber\\ 
	N_n &=& \left\{\begin{array}{ll}
		\sqrt{{q\over 2a}} (1 - e^{-qL/a})^{-1/2}(1+q^2)^{-1/4}, & \phantom{A!} n=0\\
		{1\over
		\sqrt{4L}}\left(1 + \left({q\over 2 k_n a}\right)^2 \right)^{-1/2}(1+q^2)^{-1/4},& 
		\phantom{A!} n\ge 1 \end{array} \right. \nonumber\\
	\tilde f_n(x) &=&\left\{\begin{array}{ll} 
		 {1\over \sqrt{4L}} e^{q|x|/2a}\sin(k_n x),&  \phantom{AAAAAAAAAAA!!} n\ge 1 
		 \end{array} \right.
\eeqa	
and $k_n = n\pi/L$.  

These solutions have the desirable property, from the point of view of string theory, that
there is a zero mode accompanied by a tower of heavy states \cite{MZ,HN}.  This is
qualitatively similar to the spectrum of excited states of the open string, though we have
sacrificed some of the similarity by taking the tachyon potential $e^{-|T|/a}$ instead of
$e^{-T^2/a^2}$.  With the latter choice one gets a more realistic spectrum of the form
$\bar\omega_n^2 \sim \vpp + n/a^2$, which has the correct $n$-dependence to match string
theory.  The disadvantage is that the solutions in region II cannot be found analytically
since  $\ddot A + \omega_m^2 e^{-2t^2/a^2} A=0$ is not a standard differential equation. We
have therefore given up some of the quantitative similarities with the real theory for the
sake of being able to go as far as possible analytically.

\subsection{Matching at interfaces}

To complete our task, we must relate the solutions in each neighboring region to each
other at the interfaces $t=0$ and $t=q|x|$.  This will impose relations between
$a_m,d^\dagger_m$ and $b_m,c_m$ and between $b_m,c_m$ and $d_n,d^\dagger_n$.  At $t=0$
the procedure is straightforward; $A$ and $\partial_0 A$ must be continuous, leading to
the relations
\beq
\label{bcaa}
	\left(\begin{array}{c} b_m\\ c_m \end{array} \right) = 
	{\pi a \sqrt{\omega_m}\over 2\sqrt{2}} 
	\left(\begin{array}{cc} -Y_1 - iY_0 & -Y_1 + i Y_0\\ 
	\phantom{-}J_1+iJ_0 & \phantom{-}J_1-iJ_0 \end{array} \right) 
	\left(\begin{array}{c} a_m\\ a_m^\dagger \end{array} \right)
\eeq
where the Bessel functions are all evaluated at $a\omega_m$.  We used the Wronskian of
$J_0$ and $Y_0$ to obtain (\ref{bcaa}).  Notice that there is
no mixing between different $m$ values at this point.

Matching at the $t=q|x|$ interface is more difficult.  In this case we also demand
continuity of $A$, but the fact that the prefactor $\sqrt{1-(\partial_\mu T)^2}$
in the action is discontinuous means that derivatives of $A$ are discontinuous as
well.  By integrating the partial differential equation for $A$ along an infinitesimal
path which crosses the interface perpendicularly, we can show that the following
linear combination of derivatives must be continuous at the interface:
\beq
	\Delta(F_t \dot A + q F_x A') =0 \hbox{\ from II to III}
\eeq
where $F_t$ is the coefficient of $\dot A^2$ and $F_x$ is that of $A'^2$ in the action.
Namely, $F_t = 1$ in region II and $\sqrt{1+q^2} e^{-q |x|/a}$ in III, while
$F_x = e^{-2 t/a}$ in II and $e^{-q |x|/a}/\sqrt{1+q^2}$ in III.  This gives the
rather cumbersome matching conditions
\beqa
\label{match1}
 && {1\over\sqrt{4L}} \sum_m 
   \left(b_m J_0(a\omega_m e^{-q|x|/a}) + c_m Y_0(a\omega_m e^{-q|x|/a})\right)
   \cos(k_m x)  \nonumber\\    && \phantom{AAAAAAAAAAAAAAA}
   = \sum_n {1\over\sqrt{2\bar\omega_n}}
   \left(d_n e^{-i\bar\omega_n q|x|} + d^\dagger_n e^{i\bar\omega_n q|x|}\right) f_n(x)
\eeqa
for $A$ itself and 
\beqa
\label{match2}
&& {1\over\sqrt{4L}} \sum_m \Biggl[
   \omega_m  \left(b_m J_1(a\omega_m e^{-q|x|/a}) + c_m Y_1(a\omega_m e^{-q|x|/a})\right)
   \cos(k_m x) \nonumber\\
&& \phantom{AAAAAAAAA}- q k_m e^{-q|x|/a}
   \left(b_m J_0(a\omega_m e^{-q|x|/a}) + c_m Y_0(a\omega_m e^{-q|x|/a})\right)
   \sin(k_m x) \Biggr] \nonumber\\
&&  = \sum_n {1\over\sqrt{2\bar\omega_n}}\left[
	-i\omega_n\sqrt{1+q^2}\left(d_n e^{-i\bar\omega_n q|x|} 
	- d^\dagger_n e^{i\bar\omega_n q|x|}\right)f_n(x)
	\phantom{q\over\sqrt{1+q^2}} \right. \nonumber\\ 
&& \phantom{AAAAAAAAA}	+\left. {q\over\sqrt{1+q^2}} \left(d_n e^{-i\bar\omega_n q|x|} + d^\dagger_n
	e^{i\bar\omega_n q|x|}\right) f_n'(x)\right]
\eeqa
for the derivatives of $A$.  Similar conditions hold among the odd parity modes
[$a,d\to\tilde a,\tilde d$, $f_n\to\tilde f_n$, $\cos(k_m x)\to \sin(k_m x)$], which do not
mix with those of even parity.
 
\subsection{Solution of matching conditions}
 
The technical challenge is now to solve for $d_n,d_n^\dagger$ in terms of $b_n,c_n$. 
Normally this would be done by taking the inner product of each equation 
(\ref{match1}, \ref{match2}) with some 
function which is orthogonal all but one of the functions multiplying $d_n$ or
$d_n^\dagger$.  But because of the diagonal nature of the II/III interface, the latter
functions are {\it products} of orthogonal functions, which are no longer orthogonal.
This makes it impossible to solve for $d_n,d_n^\dagger$ analytically.  

We should therefore choose some set of basis functions $g_i(x)$ and integrate them
against the matching conditions (\ref{match1}-\ref{match2}) to transform the latter into
the discrete form 
\beq
\label{ddbc}
	\sum_n L_{in} \left(\begin{array}{c} d_n\\ d_n^\dagger \end{array} \right) = 
	\sum_m R_{im} \left(\begin{array}{c} b_m\\ c_m \end{array} \right)
\eeq
We were not able to identify any set of $g_i(x)$ that makes the matrix
$L_{in}$ even approximately diagonal (which would facilitate an analytic approximation),
making a numerical solution necessary.  After some experimentation, one realizes that the
most efficient way to do this is to discretize the system on a spatial lattice at
positions $x_i$, so that $g_i(x)=\delta(x-x_i)$.  This allows us to compute the
matrices $L_{in}$ and $R_{im}$ without having to perform any integrals.  An ultraviolet
cutoff is thus introduced.  Let $i=0,\dots,N$ so that $x_i = iL/N$.  It makes sense to
let the mode number of the spatial eigenfunctions $f_n(x)$ also range from $0$ to $N$;
then (\ref{ddbc}) gives exactly as many equations as unknowns.  The system can 
be solved by numerically inverting the matrix $L_{in}$ \cite{numrec}.  
The results are presented in the next section.

\section{Particle Production}

Our solutions for the gauge field in regions I and III are normalized so that $a_m,\
a_m^\dagger,\ d_m,\ d_m^\dagger$ are the correct creation and annihilation operators for
particles in the distant  past and future once the gauge field is quantized.  We assume
the universe starts in the vacuum state $a_n |0\rangle = 0$.  But this state is not
annihilated by $d_n$, since the latter is a superposition of $a,\ a^\dagger$ determined
by the Bogoliubov coefficients $\alpha$ and $\beta$,
\beq
	\left(\begin{array}{c} d_n\\ d_n^\dagger \end{array} \right) = 
		\sum_m	\left(\begin{array}{cc} \alpha_{nm} & \beta_{nm} \\ 
	\beta^*_{nm} & \alpha^*_{nm} \end{array} \right) 
	\left(\begin{array}{c} a_m\\ a_m^\dagger \end{array} \right)
\eeq
Therefore observers in the future will see a spectrum of particles in the final state
given by
\beq
	{\cal N}_n = \langle 0| d^\dagger_n d_n | 0 \rangle = 
		\sum_m |\beta_{nm}|^2
\eeq
In this section the numerical results for ${\cal N}_n$ and the total energy density of
produced radiation will be presented. 

\subsection{UV sensitivity}

Ideally one should obtain convergent results in the limit that $N\to\infty$, where we
recall that $N$ is the number of sites of the spatial lattice in the extra dimension,
introduced  in the previous section.  However we do not observe this from our
numerical results; rather there is a steady growth with $N$.   Our conjecture is that
this is related to the discontinuity in the action which arises from the sign change in
$(\partial_\mu T)^2$ across the II/III interface.  In a simpler situation where particles
are produced due to a (spatially constant) time-varying background, the spectrum is
proportional to the square of the Fourier transform of the background.  If the latter has
sharp features, the spectrum falls only as a power of energy, whereas a smooth
background leads to exponential suppression of high energies.  For example, the
Fourier transform of  $e^{-t\Lambda}$ times a step function of time is $(\Lambda +
i\omega)^{-1}$ whereas the Fourier transform of $(1 + t^2\Lambda^2)$ is proportional to
$e^{-|\omega|/\Lambda}$. In the former case, summing over high frequency modes could lead
to nonconvergent results. We expect the high frequency contributions 
to be suppressed by a factor like $e^{-|\omega|/\Lambda}$ if we had a more realistic
tachyon profile whose derivative changed smoothly over a time $1/\Lambda$.  We have not
yet found a way of altering $T(t,x)$ to incorporate this behavior while still allowing us
to solve for $A(t,x)$ analytically. 

To remove sensitivity to the lattice spacing, we therefore try to model the expected
effects of smoothing out the background tachyon solution by inserting 
a convergence factor by hand, so that (\ref{ddbc}) becomes
\beq
\label{ddbc2}
\sum_n L_{in} \left(\begin{array}{c} d_n\\ d_n^\dagger \end{array} \right) = 
\sum_m R_{im}\, e^{-\omega_m/\Lambda} \left(\begin{array}{c} b_m\\ c_m \end{array} \right)
\eeq
Once this is done, the limit $N\to\infty$ is well-behaved.  We have thus essentially traded the
original ultraviolet cutoff $N/L$ for a new one, $\Lambda$, whose physical meaning is more
transparent.  Figure 2(a) shows the dependence on $N$ of the low-momentum zero-mode production,
${\cal N}_0(p_0)$, at $p_0=0.05/a$, for several values of $\Lambda$ and $q$ (recall that the
latter parameter determines the slope of the interface between spacetime  regions II and
III).   The convergence with $N$ is faster for smaller values of $q$; numerical limitations
therefore prevent us from accurately studying large values of $q$. Fig.\ 2(b) shows the
dependence of zero-mode production on moderate values of $q$;  large values of $q$ at fixed $N$
give exponentially increasing results due to the term $Y_1(a\omega_m e^{-qx/a})$ in
(\ref{match2}), but these  nevertheless become well-behaved again as $N$ is increased to
sufficiently large values.  On physical grounds one might anyway expect $q\le 1$ due to the
finite speed of propagation of the kink.

\subsection{Spectra}
\vskip-0.04in
We note that each state in region III is distinguished not only by its mode number $n$, but
also  its momentum $\vec p$ in the large dimensions, which we have until now
suppressed  except for its appearance in the energies $\bar\omega_n$, eq.\
(\ref{baromega}).  In figure 3 we show the dependence of ${\cal N}_n(p)$ on these two
variables. The spectrum of zero modes falls monotonically with $p$ as expected. 
Although the nonzero mode spectra temporarily rise with $p$, their
contributions to the total energy density are smaller than that of the zero mode.  We
have taken $q=1$ and $\Lambda=1/a$ for the parameters controlling the steepness the
kink in  spacetime and the suddenness with which it forms.
 The dependence on the size of the bulk $L$ can be seen in fig.\ 3(b).

Here and in the remainder we have made the simplifying approximation of
ignoring the odd-parity modes and keeping only the even ones, which include the 
dominant zero mode.  We expect that this gives a slight underestimate of the
actual efficiency of particle production, by no more than a factor of order unity.

\centerline{\epsfxsize=3.75in\epsfbox{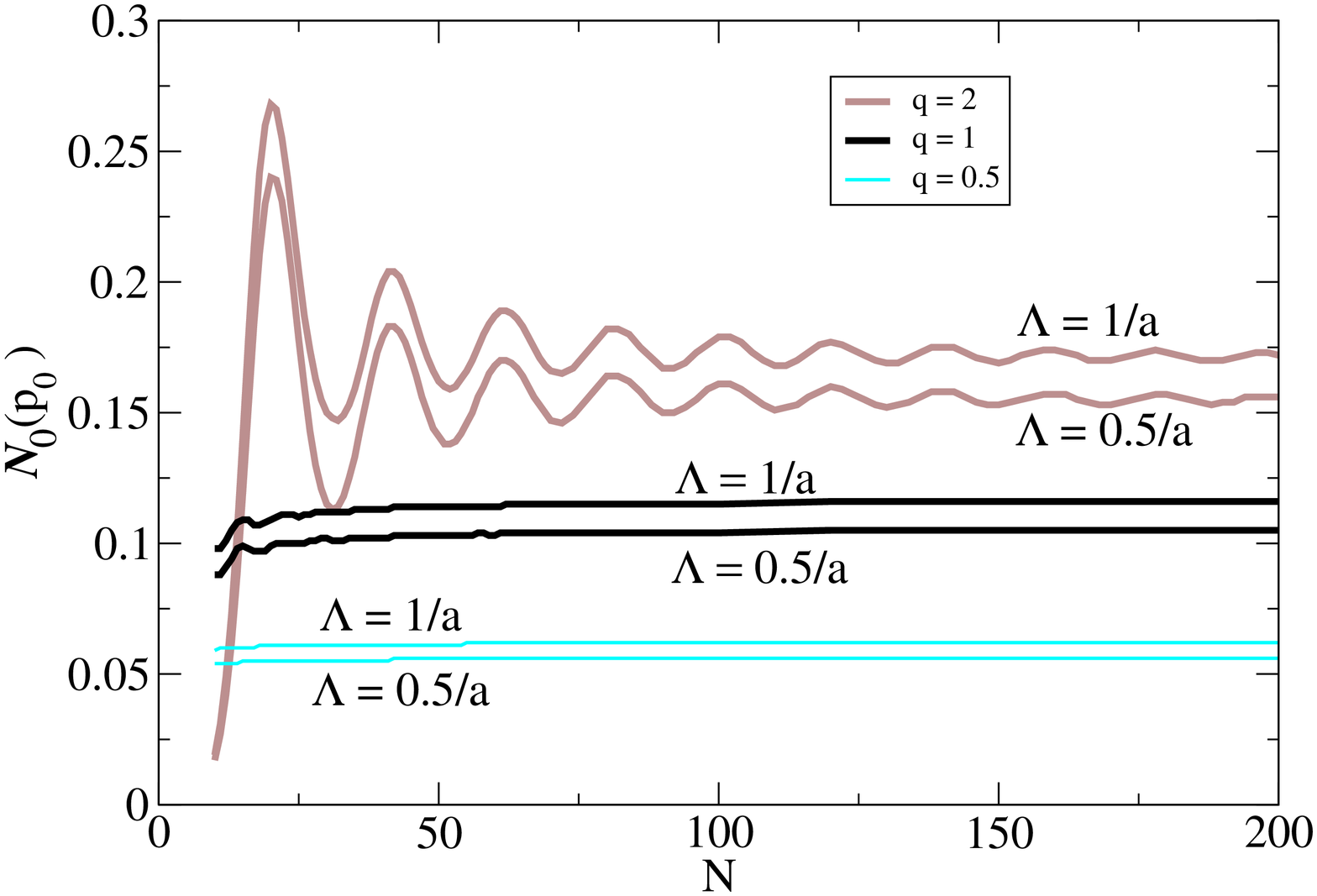}$\!\!\!\!\!\!\!\!\!\!\!\!\!\!\!\!$\epsfxsize=3.5in\epsfbox{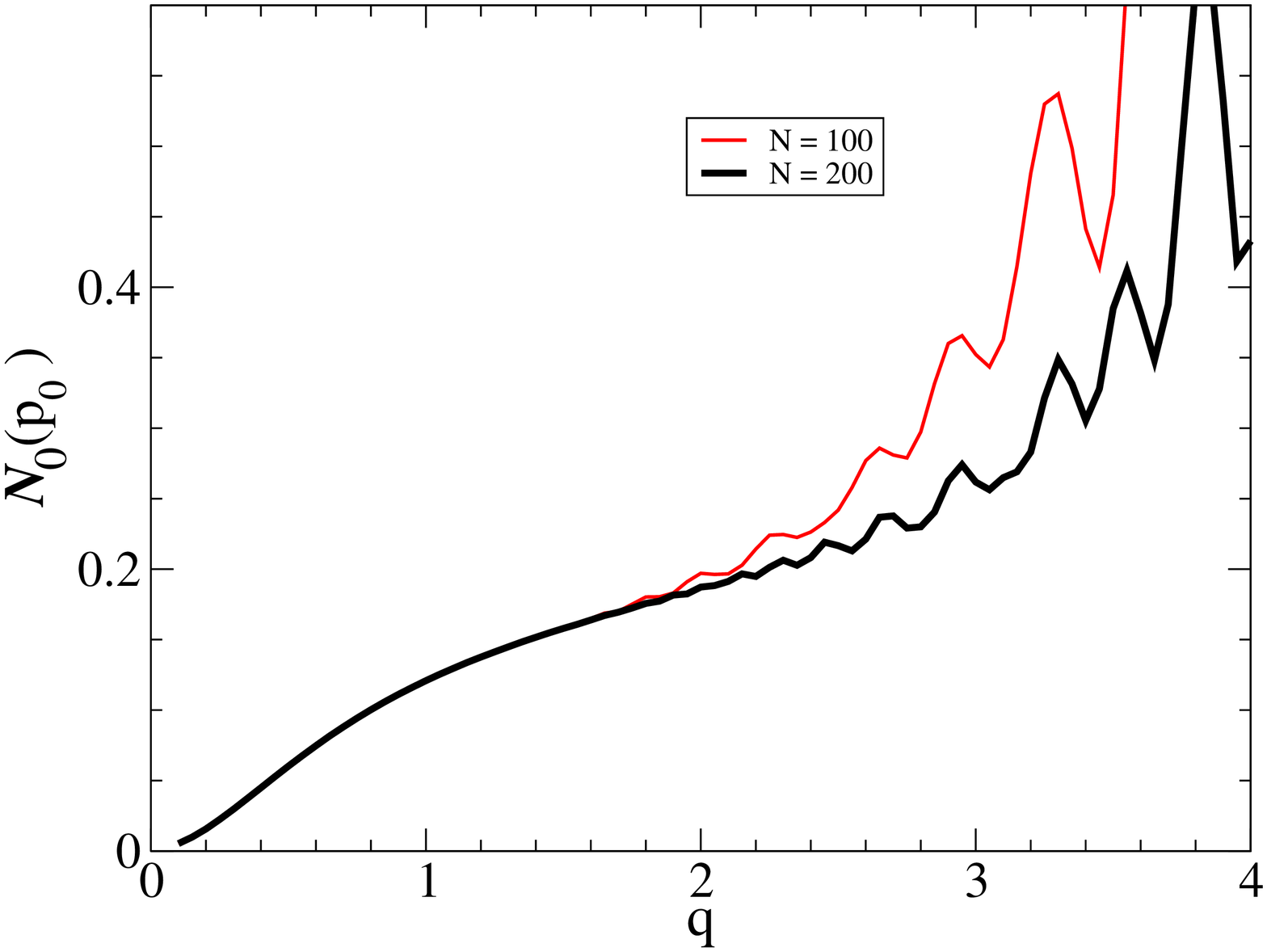}}

{\small\noindent
Figure 2. (a) Dependence of ${\cal N}_0(p_0)$ on $N$ for $p_0=0.05/a$, $L=2a$,
$\Lambda = 0.5/a,\ 1/a$ and $q=0.5,\ 1,\ 2$. 
b) Dependence of ${\cal N}_0(p_0)$ on $q$ for $N=100,\ 200$, at $p_0 = 0.01/a$, $L=2a$ and 
$\Lambda=1/a$. Physically meaningful results correspond to the $N\to\infty$ limit.}
\bigskip

To find the total energy density of produced radiation, we should sum over both $n$
and $p$:  
\beq
\label{eden}
	\rho_r = \int dp {d\rho_r\over dp} \equiv 
\sum_n \int { d^{\,3}p\over (2\pi)^3}\, {\cal N}_n(p)
\eeq
The heavier modes are counted because they will presumably decay very 
quickly into massless standard model particles.  In figure 4 we show the 
differential energy density, ${d\rho_r\over dp}$, as a function of momentum,
for several values of the other parameters. 
\vskip-0.025in
\centerline{\epsfxsize=3.5in\epsfbox{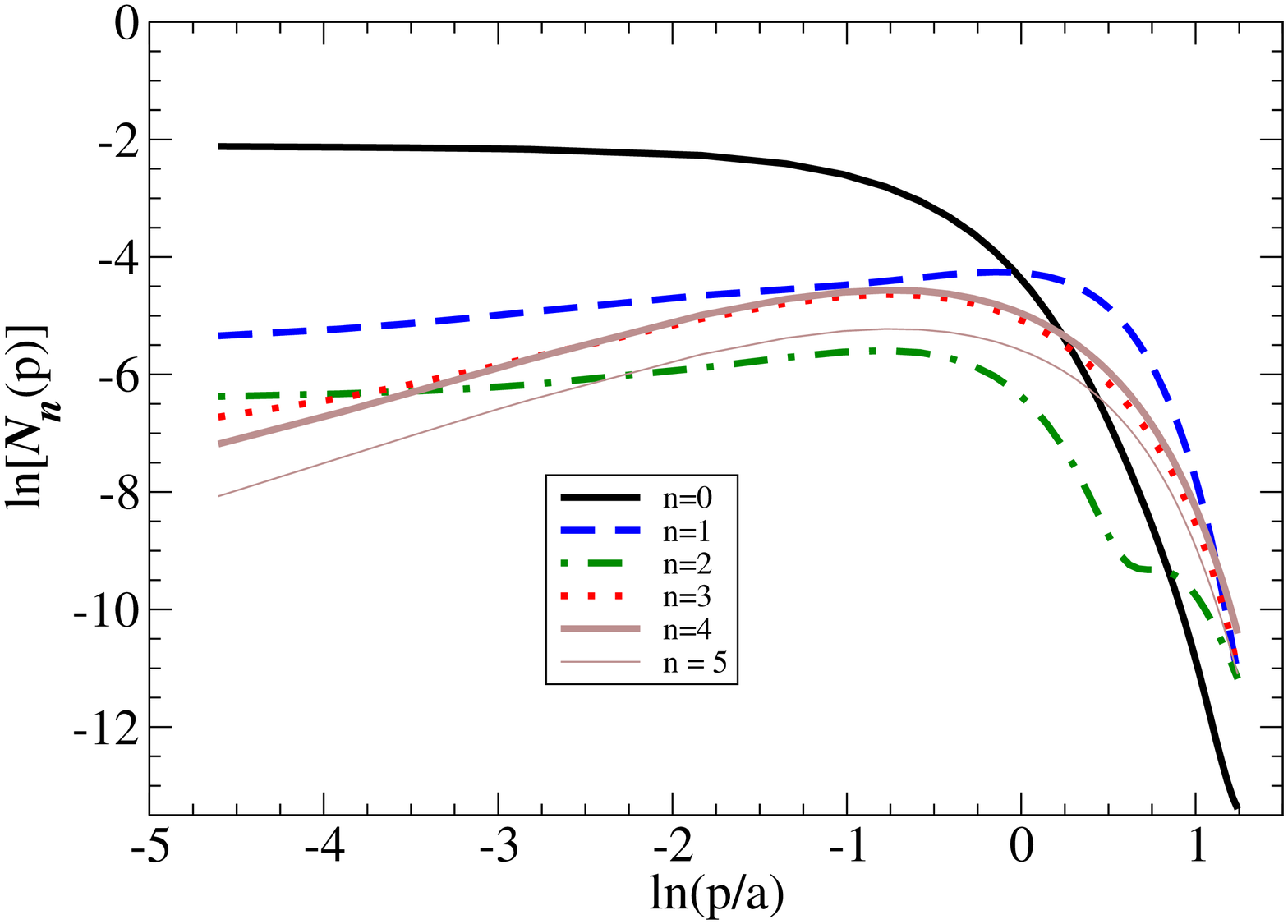}$\!\!\!\!\!\!\!\!\!\!\!\!\!$\epsfxsize=3.5in\epsfbox{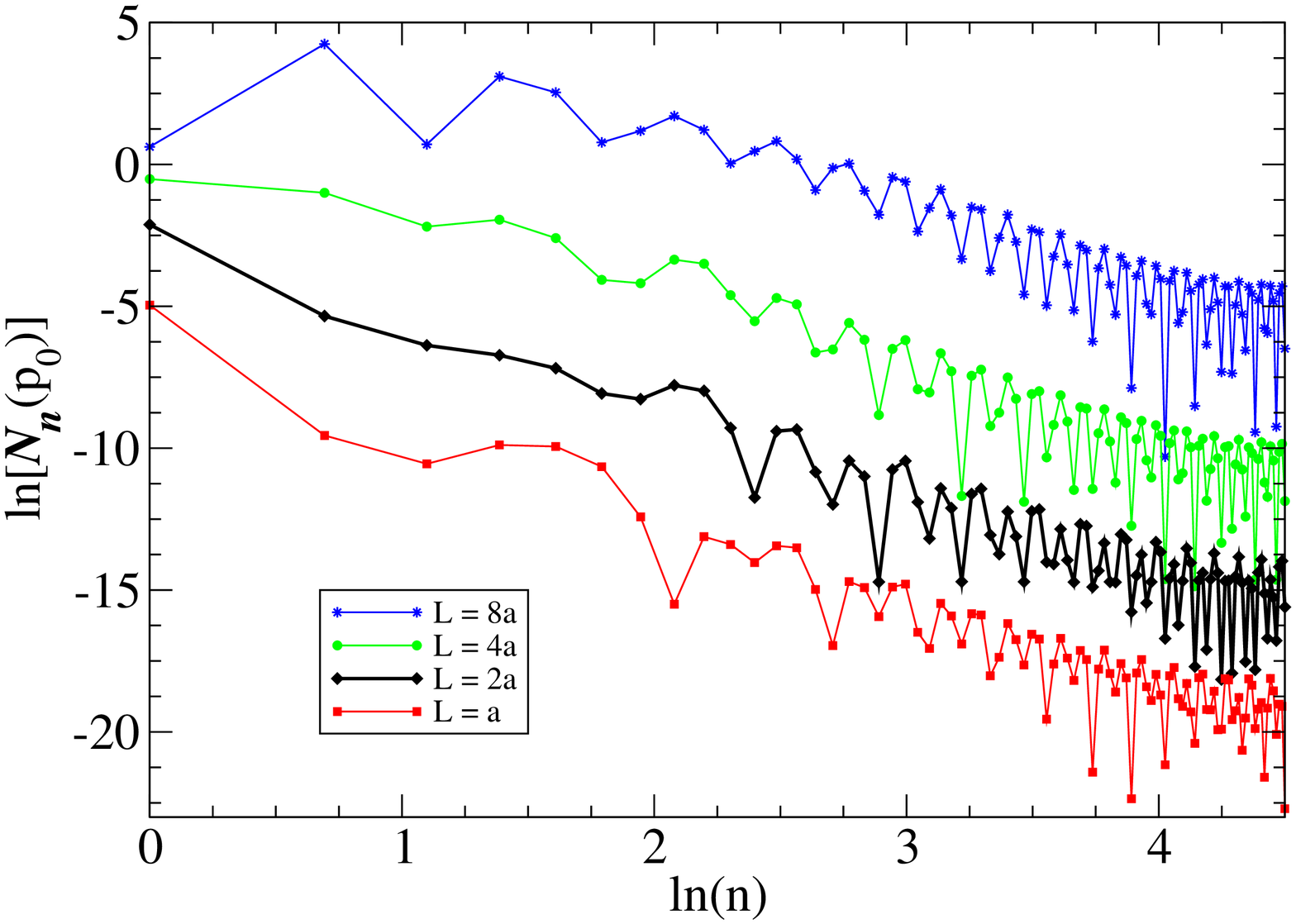}}
{\small\noindent
Figure 3.(a) Momentum dependence of the spectrum of produced particles,
${\cal N}_n(p)$ for the string excitation quantum numbers $n=0,\dots,5$.
(b) ${\cal N}_n(p)$ as a function of $n$ for $L=8a,\ 4a,\ 2a,\ a$ and $p=p_0\equiv 0.01/a$.  
Both graphs are for $q=1$, $\Lambda=1/a$ .}

\subsection{Energy density produced}
We turn now to the main results, the total energy density $\rho_r$ of produced radiation,
obtained from integrating eq.\ (\ref{eden}).  In figures 5(a-c) we graph the dimensionless 
combination $\ln(a^4 \rho_r)$ as a function of the main unknown parameters, $q$, $\Lambda$ and
$L$.  It is clearly an increasing function of all three parameters. The ``critical value''
shown in these figures, which we would like $\rho_r$ to exceed,  is derived in the next
subsection.  We will argue that this is the value at which reheating starts to 
significantly deplete the energy stored in the rolling tachyon fluid.

\vskip-0.01in
\centerline{\epsfxsize=3.75in\epsfbox{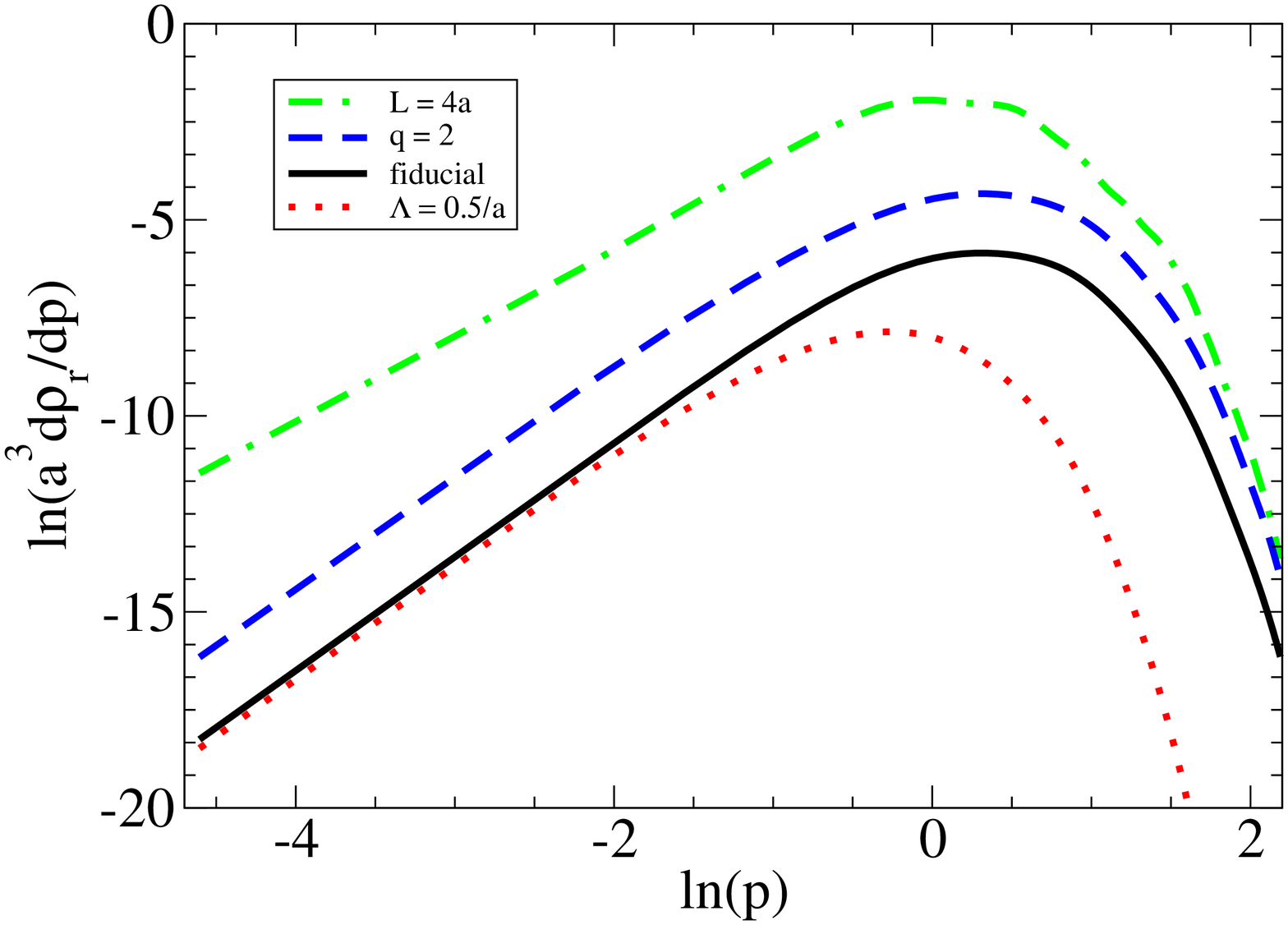}\hfil\raisebox{1.8in}{\parbox{2.in}{{\small
\noindent Figure 4. Differential energy density of produced radiation as function of
$p$.  Fiducial values of parameters are $L=a$, $\Lambda=1/a$ and $q=1$.}}}$\phantom{AAAAAAAA!!}$}
\vskip-0.15in
\centerline{\epsfxsize=3.5in\epsfbox{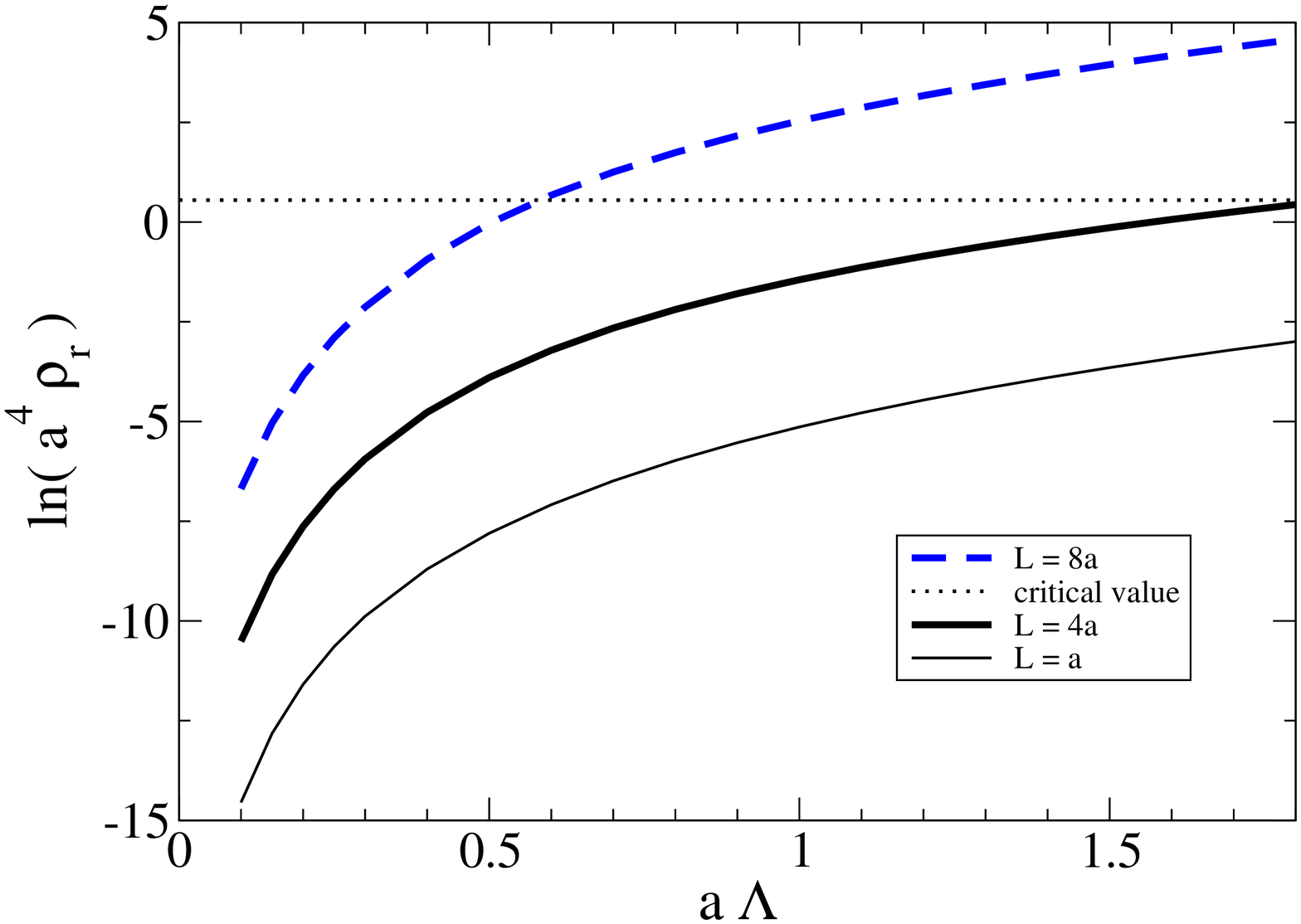}$\!\!\!\!\!\!\!\!\!\!\!\!\!$\epsfxsize=3.5in\epsfbox{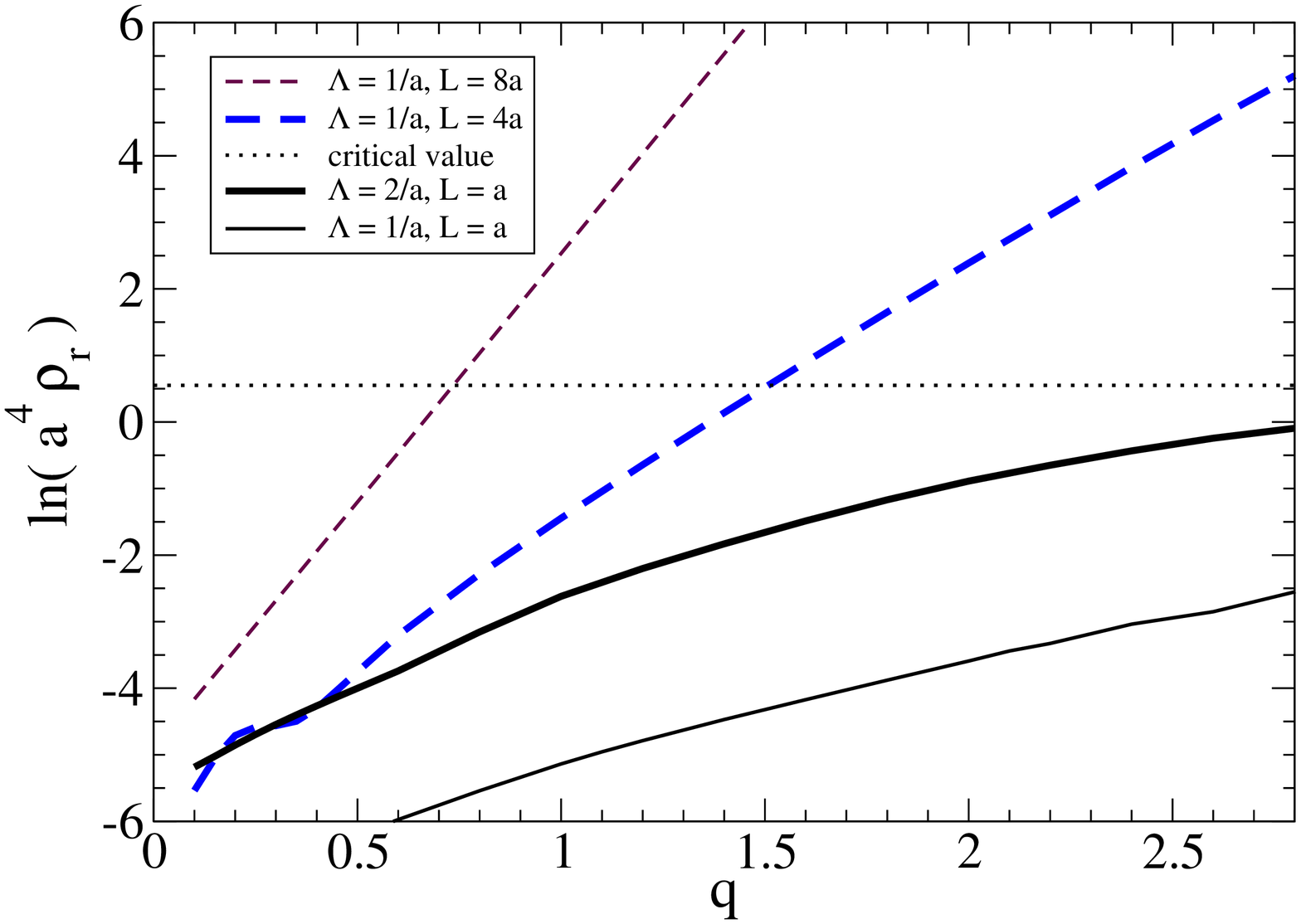}}
\vskip-0.1in
{\small\noindent
\leftline{Figure 5.(a) $\ln(a^4 \rho_r)$ versus $a\Lambda$ for $L=a,4a,8a$.  
(b) $\ln(a^4 \rho_r)$ versus $q$ for several values of $L$ and $\Lambda$.}}
\vfill\eject

\centerline{\epsfxsize=3.25in\epsfbox{fig5c.eps}\hfil\raisebox{1.8in}{\parbox{2.in}{{\small
\noindent Figure 5(c). $\ln(a^4 \rho_r)$ versus $L/a$ for  $\Lambda=2/a,\ 1/a$ and $0.5/a$,
with $q=1$.}}}}

\subsection{Efficiency of reheating}
Our goal is to see if the produced $\rho_r$
is a large enough fraction of the energy density which is available from the  
homogeneous rolling tachyon fluid, $\rho_{\sss T}$.  We thus need to specify the
value of $\rho_{\sss T}$ which is predicted by string theory.
In terms of the effective
tension of the original brane and the size of the extra dimension, 
\beq
\label{rhoT}
	\rho_{\sss T} = 2L {\cal T} 
\eeq
By the effective tension, ${\cal T}$, we mean the energy density in the $(4+1)$-dimensional
spacetime which includes $x$.  If we started from a nonBPS D$p$-brane, then ${\cal T} = 
\sqrt{2}{\cal T}_p V$ where $V$ is the volume of compact dimensions within the brane
excluding $x$, and ${\cal T}_p$ is the tension of a BPS D$p$-brane,
\beq
\label{tension}
	{\cal T}_p = {1\over g_s}(2\pi)^{-p}(\sqrt{\alpha'})^{-(p+1)}
\eeq
Therefore $\rho_{\sss T}$ depends on  $g_s$, $V$, $L$ and $\alpha'$.  However not
all of these are independent; they are related to the fine structure constant of the
gauge coupling evaluated at the string scale $M_s = (\alpha')^{-1/2}$ \cite{JST}:
\beq
	\alpha(M_s) = {g_s (2\pi\sqrt{\alpha'})^{p-3}\over 2(2LV)}
\eeq
Interestingly, if we use this to eliminate the $V$ dependence from (\ref{rhoT})
we find that the dependence on the string coupling and on $L$ is also gone:
\beq
	\rho_{\sss T} = {M_s^4 \over \sqrt{2} (2\pi)^3 \alpha(M_s)}
\eeq

To find out how much energy is available for reheating, we must also consider the descendant
brane's 4-D energy density, $\rho_f$.  This is obtained by integrating over the extra dimension
in (\ref{action3}) for the kink profile $T=qx$ and taking the limit $q\to\infty$.  The result
is simply $\rho_{f} = 2a {\cal T}$.  Therefore the excess energy density which can be used
for reheating is
\beq
\label{delrho}
	\Delta\rho = \rho_{\sss T} - \rho_f = \rho_{\sss T} \left(1 - {a\over L}\right)
\eeq
Hence we should demand that $L>a$ to get any reheating at all.

In order to compare $\Delta\rho$ to the energy density of produced radiation, we need to know
the parameter $a$ in terms of the string length.  In our simplified model of tachyon condensation,
this can be determined by demanding that the tension of the descendant $(p-1)$-brane match
the string theoretic value for a BPS state, given by (\ref{tension}) with $p\to p-1$.  In the
limit $q\to\infty$, the ratio of the initial and final brane tensions in our field theory
model is $(\lim_{q\to\infty}\sqrt{1+q^2}\int_{-\infty}^{\infty}e^{-q|x|/a} dx)^{-1} = 1/2a$,
whereas the string theoretic value is $\sqrt{2}M_s/(2\pi)$.  We therefore obtain
\beq
\label{arel}
	a = \pi \sqrt{\alpha'/2} = {\pi\over\sqrt{2} M_s}
\eeq

Now we can quantify the efficiency of reheating.  Our results for $\rho_r$ are expressed
in units of $a^{-4}$.  Using (\ref{arel}) we can convert the available energy density
$\Delta\rho$ into the same units to find the critical value of $\rho_r$, call it $\rho_c$,
for which the conversion into radiation would be 100\% efficient:
\beq
	\rho_c = {\pi\over 32 \sqrt{2}\,\alpha(M_s)}a^{-4} \cong 1.7 a^{-4}
\eeq
We take the fine structure constant to be $1/25$, and 
we omit the factor $(1-a/L)$ since it would require fine tuning of $L$ to make
it very relevant.  The figure of merit for reheating is therefore $\rho_r/\rho_c$.
The critical value $\rho_c$ appears as a dotted line in figures 5(a-c).  We see that
for large enough values of $\Lambda$, $q$, or $L$, reheating can be efficient.

\section{Conclusion}

We have made a case for reheating the universe after brane-antibrane inflation by
production of massless gauge particles, due to their coupling to the fast-rolling tachyon
field which describes the instability of the initial state.  Our results are encouraging,
indicating that if the extra dimensions within the original branes but transverse to the
final one are large enough compared to the string length scale $\sqrt{\alpha'}$, reheating
can be efficient enough so that radiation dominates over cold dark tachyon matter. 
Depending on details of the tachyon background and the compactification, an extra dimension
of size $\sim 10 \sqrt{\alpha'}$ could be sufficient.  (For example, with
$L=4a$ in an orbifolded compactification, the size of the extra dimension
would be $4\pi/\sqrt{2} M_s^{-1} \cong 9 \sqrt{\alpha'}$. 
In the present analysis we counted
only a single polarization of one $U(1)$ gauge boson and ignored the odd parity modes in
the extra dimension, so the real situation could be less constrained.

The calculation is complicated, and we have made it tractable by invoking a number of
simplifying assumptions.  We considered formation of a kink in the tachyon field rather
than a vortex.  We used a simplified version of the tachyon action rather than the one
which has been derived in BSFT.  We used an ansatz for the background tachyon field which
is a good approximation to the actual solutions in the regions where it depends only on
space (the kink) or on time (the homogeneous roll), but we do not know how suddenly it
makes the transition between these two regimes, hence we parametrized this uncertainty by
introducing a cutoff $\Lambda$ on the tachyon field's derivative.  We have also left the
slope $q$ of the tachyon kink profile $T=qx$, during kink formation, as a free parameter. 
Moreover we have ignored the expansion of the universe, which is only correct if the brunt
of the tachyon roll completes in a time not much exceeding the Hubble time.  We have also
ignored the back-reaction of the produced particles on the tachyon background, so our
criterion of requiring the produced energy density to meet or exceed that which is
initially  available is a crude one.

To do a better job, more numerical computations will probably be necessary since it is
hard to solve the gauge field equations of motion in a tachyon background that depends on
both space and time.  One check that might be carried out relatively easily is to
numerically obtain the time- and space-dependent solution $T(x,t)$ for the more 
realistic action and compare its features to those we have assumed.  This is in progress.
Very recent work on how to generate time-dependent tachyon solutions in ref.\ 
\cite{sen-timedep} might prove to be helpful here.

As we were finishing this paper, a related work appeared \cite{STW}.  The latter assumes
that there is a constant rate of decay of the tachyon fluid into radiation.  We have
presented arguments to the contrary.  If such an assumption could be justified, it would
greatly enhance the viability of inflation via brane-antibrane annihilation.

\bigskip

We thank C.\ Burgess for useful suggestions during the course of this work.
We also thank D.\ Easson, K.\ Bardakci, A.\ Sen, P.\ Kraus, P.\ Steinhardt, and H.\ Tye 
for helpful comments and correspondence.

\end{document}